\documentclass[doublecol]{epl2}

\usepackage{graphicx}
\usepackage{amsmath}
\usepackage{amsfonts}
\usepackage{amsthm}
\usepackage{amssymb}
\usepackage{bm}
\newcommand{\be}{\begin{equation}}
\newcommand{\ee}{\end{equation}}
\newcommand{\kk}{\bm{k}}

\newcommand{\ham}{{\cal H}}
\newcommand{\bg}{{\bf g}}
\newcommand{\dkx}{{\partial_{k_x}}}
\newcommand{\dky}{{\partial_{k_y}}}

\title{Topological Weyl Semi-metal from a Lattice Model}

\author{Pierre Delplace\inst{1} \and Jian Li\inst{1} \and David Carpentier \inst{2}}
\institute{
\inst{1} D{\'e}partement de Physique Th{\'e}orique, Universit{\'e} de Gen{\`e}ve, CH-1211 Gen{\`e}ve, Switzerland \\
\inst{2} CNRS - Laboratoire de Physique, Ecole Normale Sup{\'e}rieure de Lyon, 46,
All{\'e}e d'Italie, 69007 Lyon, France
}


\abstract{
We define and study a three dimensional lattice model which displays a Weyl semi-metallic phase.
This model consists of coupled layers of quantum (anomalous) Hall insulators. The Weyl semi-metallic phase appears between
 a resulting quantum Hall insulating phase and a normal insulating phase. Weyl fermions in this Weyl semi-metal, similar 
to Dirac fermions in graphene, have their lattice pseudo-spin locked to their momenta. We investigate surface states and Fermi arcs, and their evolution for different phases, by exactly diagonalizing the lattice model as well as by analyzing their topological origins.
}

\begin{document}
%\title{Surface bands and Fermi arcs in Weyl semi-metals}
%
%\author{US}
%\affiliation{}
%
%\date{\today}
%
%\begin{abstract}
%Abstract
%\end{abstract}
%

\maketitle

%%%%%%%%%%%%%%%%%%%%%%%%%%%%%%%%%%%%%%%%%%%%%%%%%%
\section{Introduction}

The interests in topological phases of condensed matter have been completely renewed by the discovery of the $\mathbb{Z}_{2}$ topological order,
 first in two dimensions (2D) \cite{Kane:2005,bernevig06,konig07} and soon after in three dimensions (3D) \cite{Fu:2007,Moore:2007,Roy:2009}. 
This type of order is not described by Landau's symmetry breaking theory, but associated with topological properties of electronic structures \cite{Hasan:2010,Qi:2011}.
 More recently, following pioneering work in Helium 3 \cite{Volovik:2003,Volovik:2007}, the focus has shifted to topological order in semi-metallic phases. 
Such a topological semi-metal has been predicted to occur in pyrochlore iridates as a result of a combination of strong spin-orbit coupling, 
Coulomb interaction and an antiferromagnetic order preserving inversion symmetry \cite{Wan:2011}. More generally, it has been argued that a semi-metallic
 phase appears in 3D between a normal insulator and a $\mathbb{Z}_{2}$ topological insulator when inversion symmetry is broken \cite{Murakami:2007}.
 Proposals following this line include either superlattices composed of layers of normal and topological insulators subject to time-reversal or inversion 
symmetry breaking \cite{Burkov:2011,Burkov:2011b,Zyuzin:2012}, or magnetically doped bulk Bi$_{2}$Se$_{3}$\cite{Cho:2011}. These 3D semi-metals, 
or Weyl semi-metals (WSMs), possess gapless chiral surface states and open Fermi arcs terminating at the projections of the Weyl points \cite{Burkov:2011,Wan:2011}.
Such unusual features can be manifested through {\it e.g.} non-zero Hall conductivity  \cite{Wan:2011,Burkov:2011,Hosur:2012}. 
Recently, realization of nodal semi-metals where bands cross along lines has been proposed \cite{Burkov:2011b} as well as topological semi-metals in 
fermionic optical lattices \cite{Sun:2012}.

In 3D WSMs, the valence and the conduction bands touch at points (termed \textit{Weyl points}) where the dispersion relation is linear, in analogy to graphene as a 2D (Dirac)
 semi-metal \cite{CastroNeto:2009}. The simplest Hamiltonian describing the linear dispersion around such a point $W$ is the Weyl Hamiltonian ${\ham}_W({\bm{q}}) = v_{ij}q_{i}\sigma_{j}$ ($i,j=1,2,3$),
 spanned in terms of the $2\times 2$ Pauli matrices $\sigma_{i}$, and characterized by a topological Chern number (that we will call \textit{helicity} by extension of the graphene case)
 $n_{W}=\textrm{sgn}\left[\det (v_{ij})\right]=\pm 1$ \cite{Young:2011}. The Nielsen-Ninomiya theorem imposes that Weyl points must occur in pair(s)
 with opposite helicity, in any lattice model \cite{Nielsen:1981}. When inversion symmetry is present, two Weyl points $W_{\pm}$ located at opposite positions 
in the momentum space have opposite helicity $n_{W_{-}} = -n_{W_{+}}$. On the other hand, when time-reversal symmetry is present, two Weyl points of
opposite momenta have the same helicity $n_{W_{-}} = n_{W_{+}}$ if the $\sigma$ matrices describe real spin. Therefore, to reconcile the presence of both symmetries, two pairs of Weyl points must coexist.
%:at each momentum ($\mathbf{W}_+$ and $\mathbf{W}_-=-\mathbf{W}_+$), two Weyl points from different pairs coexist and have opposite helicity.
Such a situation is described by the Dirac Hamiltonian spanned in terms of the $4\times 4$ $\gamma$ matrices. 
The conditions for protection of this unstable 3D Dirac semi-metal by crystallographic symmetries have been worked out in ref.~\cite{Young:2011}.
Nevertheless, generic perturbations coupling the two pairs of coexisting Weyl points open a gap -- to recover a robust 3D semi-metal, either inversion or time-reversal symmetry needs to be broken.

In this paper, we construct a simple lattice model with broken time-reversal symmetry and preserved inversion symmetry. 
This model considers spinless fermions on a 3D lattice with properly chosen magnetic flux, and gives rise to a topological WSM phase in its phase diagram. 
Similar to the case of graphene, the spinors in our model correspond to pseudo-spin that originates from the presence of two sub-lattices. 
The Weyl points are obtained without resorting to interaction, spin-orbit coupling or magnetic order as in the case of real spin \cite{Wan:2011}.
While recent proposals for WSMs were based on perturbing 3D $\mathbb{Z}_{2}$ topological insulators, the WSM in our model is found between a $d=2+1$ 
quantum Hall insulator (QHI) and a normal insulator (NI). The $d=2+1$ QHI was initially proposed for layered quantum Hall systems \cite{Balents:1996,Chalker:1995}, 
and is obtained similarly in our model by layering 2D quantum (anomalous) Hall insulators. The gapless chiral surface states of the $d=2+1$ QHI lead to a closed
 Fermi surface that is smoothly evolved from the open Fermi arc of the WSM through phase transition. We expect this transition to manifest itself in the 
measurement of Hall conductivity. In the following of this paper, we start by introducing our model and investigating its phase diagram. Owing the simplicity 
of this lattice model, we then examine explicitly the surface bands as well as the resulting Fermi arcs in different phases. We provide afterwards a topological 
analysis which accounts for both the bulk and the surface properties. Finally we comment on the evolution of the Fermi arcs with respect to the global phase diagram.

%%%%%%%%%%%%%%%%%%%%%%%%%%%%%%%%%%%%%%%%%%%%%%%%%%
\section{Model and Phases}
\begin{figure}
    \centering
    \includegraphics[width=0.42\textwidth]{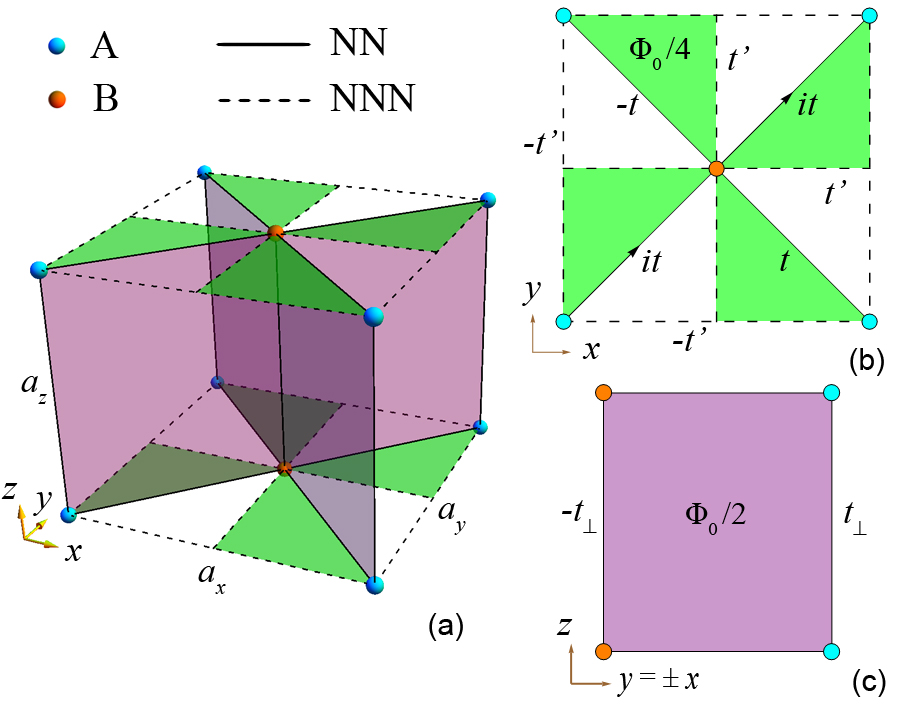}
    \caption{Illustrations of the lattice model with (a) the 3D unit cell, (b) the 2D unit cell in the $x$-$y$ plane,
and (c) part of the unit cell in the vertical plane which contains both sublattices. 
Nearest-neighbor hopping and next-nearest-neighbor hopping are represented respectively by solid and broken lines. 
Each Colored region in the unit cell is threaded by a nonzero magnetic flux: $\Phi_0/4$ out of plane towards the reader for green,
 and $\Phi_0/2$ for purple, with $\Phi_0=h/e$ being the magnetic flux quantum. One specific choice of gauge which is used in this paper is also indicated in (b) and (c).}
    \label{fig:lattice}
\end{figure}
We consider  a tight-binding model of spinless fermions defined on a 3D lattice constituted  of layers of face-centered square lattice with sub-lattices labeled by $A$ and $B$
(see Fig.~\ref{fig:lattice}). The nearest neighbor hopping between $A$ and $B$ sub-lattices is denoted by $t$, and the second-nearest neighbor hopping between $A$-$A$ or
 $B$-$B$ sub-lattice sites is $t'$.  We also introduce an on-site energy difference $2\Delta$ between the two sublattices. The planar lattices are coupled with the (nearest neighbor)
 inter-layer hopping $t_\perp$ between the $A$-$A$ or $B$-$B$ sites of adjacent layers.

The WSM phase arises when certain magnetic flux patterns are applied to this lattice. For simplicity, we will focus in this paper on one of these patterns, indicated in Fig. \ref{fig:lattice}. There are two noteworthy properties of this flux pattern: it preserves the point group symmetry $C_{4h}$ and therefore inversion symmetry, and the flux through each surface of the unit cell is zero modulo the flux quantum $\Phi_0=h/e$. As a consequence of the second property, the magnetic Brillouin zone and the original Brillouin zone of the lattice overlap exactly, therefore we will not distinguish them in the following. The first property simplifies our discussions about surface bands by relating {\it e.g.} surfaces parallel to the $x$-$z$ plane with those parallel to the $y$-$z$ plane.
%Moreover the reflection symmetry with respect to the $x$-$y$ plane that Weyl points exist,  in pairs
%(except at exactly the fusion points) and each pair related by the reflection symmetry will have opposite Chern number, or helicity -- we will come back to this point in more details
%later. The other important fact is that although the local magnetic flux breaks time-reversal symmetry,
%\textit{the total flux through any face of the unit cell is identically
%zero modulo the magnetic flux quantum $\Phi_0=h/e$}. As a consequence the magnetic Brillouin zone and the original Brillouin zone of the lattice overlap exactly,
%therefore we will not distinguish the two in our following discussions.

With minimal coupling, the effect of the magnetic flux is to decorate the hopping terms with additional phase factors that are compatible with the flux pattern \cite{Haldane:1988}. This permits various gauge choices but different choices do not leads to different physics \footnote{Note that
the inversion symmetry, which can be represented by an operator $I=\sigma_z$ for our Hamiltonian (\ref{eq:ham3D}), is preserved by an choice of gauge, {\it i.e.} it satisfies
 $I\ham(\kk)I^{-1}=\ham(-\kk)$.}. For the specific choice of gauge indicated in Fig. \ref{fig:lattice}b and c, the Bloch Hamiltonian for the bulk of the lattice reads
\begin{multline}
\hspace{-4mm}\ham(\kk) = (2t\sin \kk\cdot\bm{a}_+) \sigma_x + (2t\sin \kk\cdot\bm{a}_-) \sigma_y %\nonumber
\\
\hspace{-4mm} + \left[\Delta-2t'(\cos\kk\cdot\bm{a}_x + \cos \kk\cdot\bm{a}_y) + 2t_\perp \cos \kk\cdot\bm{a}_z \right]\sigma_z,
\label{eq:ham3D}
\end{multline}
where $\bm{a}_x$, $\bm{a}_y$ and $\bm{a}_z$ are the lattice vectors, and $\bm{a}_{\pm} = (\bm{a}_x \pm \bm{a}_y)/2$. For simplicity we consider $a_x = a_y = a_z = 1$ in the rest of this paper.
\begin{figure}
    \centering
    \includegraphics[width=0.45\textwidth]{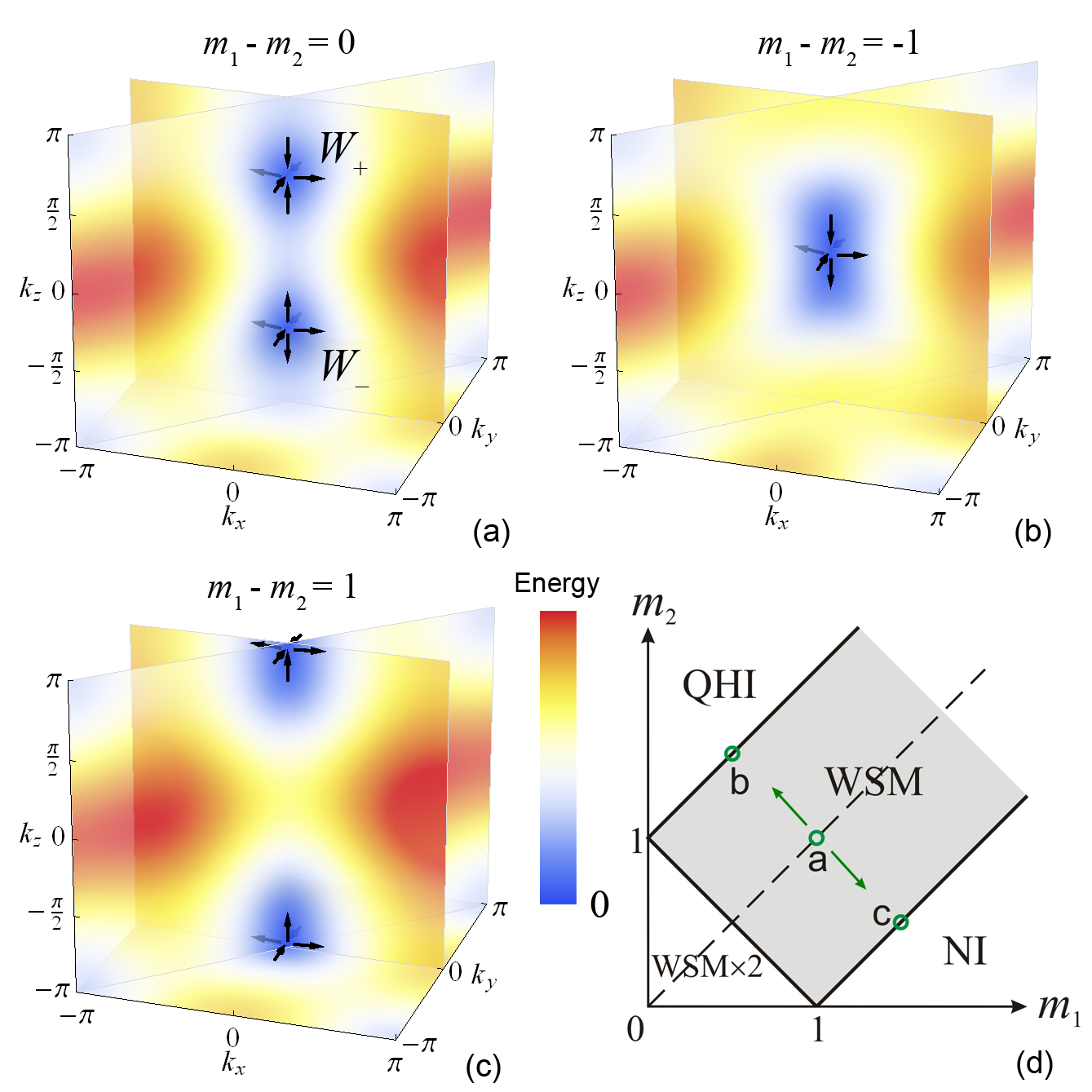}
    \caption{Color-coded 3D bulk band structures of (a) an ideal Weyl semi-metal (WSM), (b) and (c) semi-metals
when the Weyl points merge at $\bold{G}/2=(0,0,0)$ and $(0,0,\pi)$, respectively.
 All the three cases are marked in the phase diagram of the model, plotted in (d). 
The two different Weyl-point-merging processes represented by (b) and (c) lead to two topologically distinct insulating phases after the bulk gap is opened: 
one features gapless chiral surface states and is a $d=2+1$ quantum Hall insulator (QHI); the other is a normal insulator (NI). In the WSM$\times$2 phase, two pairs of Weyl points coexist.}
    \label{fig:band}
\end{figure}
When $t_\perp \ne 0$, the two 3D bulk bands are given by
\begin{align}\label{eq:E3D}
E(\kk) = & \pm 2\Bigl[t^2(\sin^2{k_+}+\sin^2{k_-}) + \nonumber\\
& t_\perp^2(m_1 - m_2\cos{k_+}\cos{k_-} + \cos{k_z})^2\Bigr]^{1/2},
\end{align}
where $k_{\pm}=(k_x \pm k_y)/2$, and we have defined two dimensionless masses $m_1=\Delta/2t_\perp$, $m_2=2t'/t_\perp$.
Without loss of generality
we will focus for now on the range $m_1 \ge 0$ and $m_2 \ge 0$, and discuss the full range at the end of the paper.

The 3D bulk bands are gapped except at points where $\sin{k_+}=\sin{k_-}=0$ and $\cos{k_z}=-(m_1 - m_2\cos{k_+}\cos{k_-})$.
The first (two) conditions requires $(k_x,k_y) = (0,0)$ or $(\pi,\pi)$. Consequently the last condition reduces to
\begin{subequations}
\label{eq:zcond}
\begin{align}
&\cos{k_z}=-(m_1 - m_2), \quad \mbox{for}\;(k_x,k_y) = (0,0); \label{eq:zcond0}\\
&\cos{k_z}=-(m_1 + m_2), \quad \mbox{for}\;(k_x,k_y) = (\pi,\pi). \label{eq:zcondpi}
\end{align}
\end{subequations}
Each of the above two equations, when fulfilled with double-valued $k_z \in [-\pi,\pi)$, gives rise to two Weyl points related by inversion symmetry and hence having opposite helicities (or Chern numbers).

Let us first suppress the pair at $(k_x,k_y) = (\pi,\pi)$ by setting $m_1 + m_2>1$ and validate the other at $(k_x,k_y) = (0,0)$
by setting $|m_1 - m_2|<1$. This is the WSM phase regime (see Fig. \ref{fig:band}d) where the bulk bands only touch at two distinct
 points in the momentum space, denoted by $W_{\pm} = (0,0,\pm\arccos{(m_2-m_1)})$. To examine these two points more closely, we
 expand the bulk Hamiltonian \eqref{eq:ham3D} around them. In the particular case when $m_1-m_2 =0$, the expansions up to linear
order in $\kk$ for $W_{\pm} = (0,0,\pm\frac{\pi}{2})$ and
after a global unitary (gauge) transformation $\ham \rightarrow U^\dagger\ham U$
with $U = \exp({-i\frac{\pi}{8}\sigma_z})$ for the purpose of clarity, lead to
%and $\tilde{\ham}(\kk) = U^\dagger \ham(\kk) U = (2\sqrt{2}t\sin{k_x/2}\cos{k_y/2})\sigma_x - %(2\sqrt{2}t\cos{k_x/2}\sin{k_y/2})\sigma_y + 2t_\perp(m_1 - m_2\cos{k_+}\cos{k_-} + \cos{k_z})\sigma_z$,
\begin{align}
  \ham_{\pm}(\bm{q}) = \sqrt{2}t~ q_x \sigma_x - \sqrt{2}t ~q_y \sigma_y \mp 2t_\perp q_z \sigma_z,
\label{eq:hamW}
\end{align}
where $\bm{q} = \kk - W_{\pm}$ for the two points respectively. These are the simplest
defining Hamiltonians for Weyl fermions, from which
one can immediately identify the helicity associated with $W_+$ to be $+1$ and with $W_-$ to be $-1$
(see Fig. \ref{fig:band}a). As $|m_1-m_2|$ is increased from $0$,
the two Weyl points move closer to each other in the momentum space, and quadratic
terms in $k_z$ in the Hamiltonian expansions become more relevant.
However, the helicity associated with each Weyl point $W_{\pm}$ remains quantized and
unchanged due to its topological nature.

The inversion symmetry %which transforms $\bold{k}$ in $-\bold{k}$,
forces the merging of a pair of Weyl points to occur
at $\bold{G}/2$ where $\bold{G}$ is a reciprocal lattice vector.
This defines the possible boundaries between the WSM phase and the bulk insulating phases. Within our model the boundaries are given by either $m_1 - m_2 = - 1$ (the upper boundary) or $m_1 - m_2 = + 1$ (the lower boundary).
There, the two bulk bands touch at only one single point with vanishing helicity:
%(it is still assumed that $m_1 + m_2>1$)
 $(0,0,0)$ for the upper boundary
(see Fig. \ref{fig:band}b), and $(0,0,\pi)$ for the lower boundary (see Fig.~\ref{fig:band}c). Compared with \eqref{eq:hamW},
the expansion of the Hamiltonian around the single touch point contains $k_z^2\sigma_z$ as the leading order term in $k_z$ (up to a constant factor), in agreement with previous studies of merging of Weyl points \cite{Murakami:2007} or 2D Dirac points \cite{Montambaux:2009}.
%In addition, the associated helicity is identically zero since such a touch point corresponds to the fusion of two Weyl points of opposite polarity.
%
%What is more interesting here is that,
%depending on at which boundary the two Weyl points are fused, the resulting insulating phases are topologically different in nature.
Beyond these transition lines two (bulk) insulating phases are reached, denoted by QHI and NI in figure \ref{fig:band}d. These two gapped phases are topologically distinct, as we will demonstrate below by first examining the surface bands and their associated Fermi arcs, and then analyzing the topological origin of the surface bands.

\section{Surface Bands and Fermi Arcs}
An advantage of a simple tight-binding lattice model is that it facilitates
explicit investigations of surface bands and Fermi arcs in the  WSM and its related insulating phases.
 As a consequence of the layered structure of our model, a strong anisotropy is present :
%An important fact one would first notice is the strong anisotropy in terms of the presence of surface bands:
no surface bands exist for surfaces perpendicular to the $z$ axis, while for surfaces parallel to the $z$ axis we find generically in the WSM phase
surface bands    showing consistent chirality with respect to the $z$-axis.
%This anisotropy is clearly inherited from the anisotropy of the lattice symmetry. More fundamentally,
%
\begin{figure}
    \centering
    \includegraphics[width=0.4\textwidth]{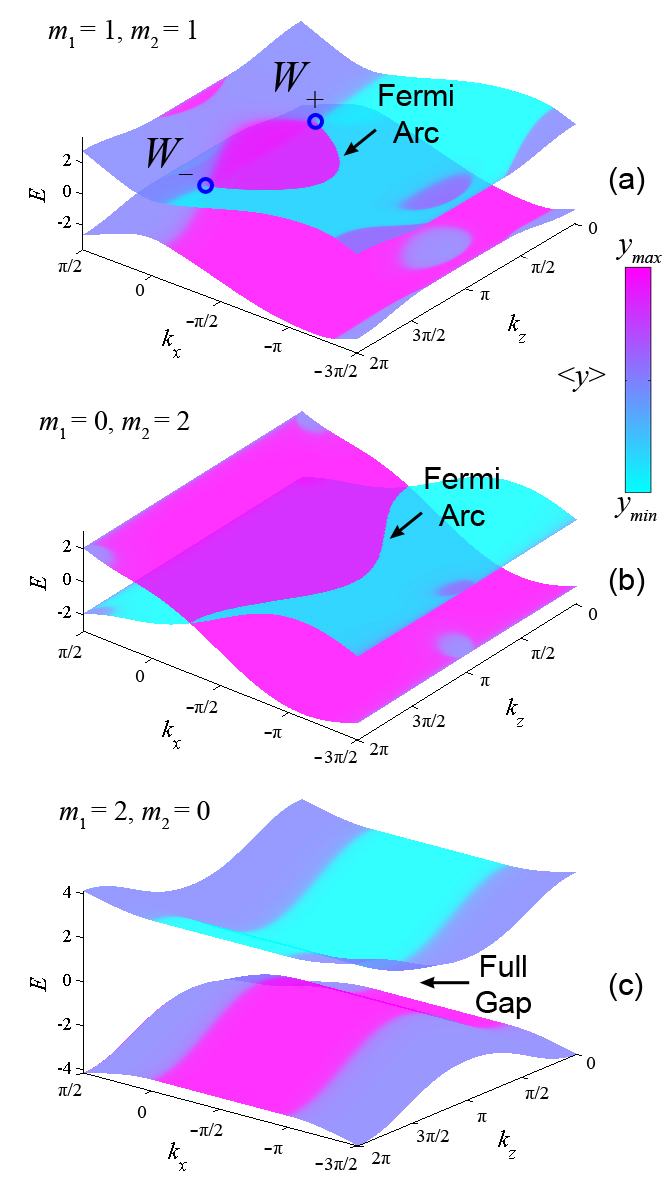}
    \caption{Surface bands for three different phases (represented by three combinations of $m_1$ and $m_2$) corresponding to (a) a Weyl semi-metal (WSM), 
(b) a $d=2+1$ quantum Hall insulator (QHI), and (c) a normal insulator (NI). Explicitly, we plot the two lowest-energy bands for samples built from our
 lattice model that are finite in $y$ but infinite in $x$ and $z$. The color code shows the average position $\langle y\rangle$ of each state, 
such that states localized at either surface ($\langle y\rangle \approx y_{min}$ or $y_{max}$) or extended in the bulk ($\langle y\rangle \approx (y_{min}+y_{max})/2$) 
can be clearly identified. Gapless chiral surface bands are present in the WSM and the QHI, but absent in the NI. The Fermi surface (assuming the Fermi energy to be zero) 
for the WSM appears to be open arcs terminating at the two Weyl points, and for the QHI appears to be closed lines across the surface Brillouin zone.
 The three scenarios here can be related by examining the evolution of surface bands and Fermi arcs upon merging of Weyl points.}
    \label{fig:surf}
\end{figure}
Here we present as examples, in Fig. \ref{fig:surf}, the surface bands for two opposite surfaces parallel to the $x$-$z$ plane
 (such that $k_x$ and $k_z$ are good quantum numbers), for the three different phases : the WSM and
% when the system is in the WSM regime, or one of
 the two bulk insulating regimes.
 As mentioned previously, the results for surfaces parallel to the $y$-$z$ plane can be
inferred by using the four-fold rotational symmetry that is inherent to our model.

In the WSM case (see Fig.~\ref{fig:surf}a; we set $m_1=m_2=1$ without losing generality), the surface bands are gapless with
respect to varying $k_x$ for a specific $k_z \in [\pi/2,3\pi/2]$,\footnote{Note that we have shifted the Brillouin zone from
what is used in the previous discussions for clarity of presentation.} and are fully gapped for $k_z$ in the complementary range.
In other words, for a fixed $k_z$ in the first range surface states sprout from one bulk band (conduction or valence band) and merge into the other one as $k_x$ is varied; otherwise for $k_z$ in the second range surface states evolve between two parts of the same bulk band. The transition between the two types of behavior occurs at the two
Weyl points projected to the surface Brillouin zone, $(k_x,k_z)= (0,\pi/2)$ and $(k_x,k_z)= (0,3\pi/2)$, where the conduction band and the valence band touch each other.
Surface states merge into the bulk bands exactly at these touch points. An immediate consequence is the
existence of a Fermi arc (for each surface band) connecting the two projected Weyl points. More generally, this observation is valid as long as the surface is not perpendicular to the $z$ axis such that the projected Weyl points do not overlap. The presence of Fermi arcs is a robust feature that indeed has a deep
topological root \cite{Wan:2011}. However we emphasize that the specific shape of a Fermi arc is determined by the
dispersion relation of the attached surface band, which is in turn
 determined by the boundary condition associated with the specific surface. Therefore by deforming one surface, or simply by
ending the crystal with different sublattices, the shape of the Fermi arc may change drastically even when the same surface orientation is maintained.

When the two Weyl points approach each other, the band structure deforms continuously, and the Fermi arcs
of the gapless surface states persist in a certain range of the surface Brillouin zone.
Depending on whether the two Weyl points approach each other through the Brillouin zone center or the Brillouin zone boundary, the Fermi arcs can either shrink or stretch.
It follows that, after the Weyl points merge and the bulk gap opens, the Fermi arcs either disappear (see Fig. ~\ref{fig:surf}c) or persist across the surface Brillouin zone (see Fig. ~\ref{fig:surf}b). These two different scenarios for the surface states correspond to two different phase transitions for the bulk (see Fig.~\ref{fig:band}d). The first case corresponds to a transition from the WSM to the NI phase, the latter being topologically trivial and accommodating no surface states. The second case, on the other hand, corresponds to a transition from the WSM to the $d=2+1$ QHI phase \cite{Balents:1996,Chalker:1995} where chiral surface states persist while the bulk gap opens -- this QHI phase can be naively expected by layering the standard integer quantum (anomalous) Hall insulators. Hence in this model, we find the WSM to occur between a normal and a topological (quantum Hall) insulator.

\section{Topological Order}
We now turn to an in-depth discussion about the origin of these surface states and their relation to the Weyl points.
Our discussion will be based on the dimensional reduction method which involves treating one or more momenta,
when they are good quantum numbers, as parameters of the Hamiltonian.%, and thus reducing the system effectively to
%a lower dimension that is more familiar to us or easier to study.
%
The natural starting point for this layered model is to treat $k_z$ as an effective parameter and reduce the system effectively to the 2D $x$-$y$ plane.
This reduction is valid for the bulk system or any surface parallel to the $z$-axis.
For convenience and with a minor abuse of notation, we define a $k_z$-dependent
(dimensionless) mass $m_1(k_z) = m_1 + \cos{k_z}$, and rewrite Hamiltonian \eqref{eq:ham3D} as
\begin{align}\label{eq:ham2D}
&\ham(\kk) = \bg\cdot\bm{\sigma} \\
\mbox{with}\quad &\mathrm{g}_x = 2t\sin k_+,\; \mathrm{g}_y = 2t\sin k_-, \nonumber\\
& \mathrm{g}_z = 2t_\perp\left[m_1(k_z) - m_2\cos{k_+}\cos{k_-}\right], \nonumber
\end{align}
where $\bm{\sigma}=(\sigma_x,\sigma_y,\sigma_z)$.
This 2D model is essentially equivalent to the Haldane model for Landau-level-free quantum Hall effect \cite{Haldane:1988}.
The 2D bulk bands, when fully gapped ($m_1(k_z) \mp m_2 \ne 0$), can be characterized by using the
topological (Chern) invariant defined in the momentum space as
\begin{align}\label{eq:cnum}
c(k_z) = \frac{1}{4\pi}\int_{BZ} {dk_x dk_y} \:\hat{\bg} \cdot \left( \dkx \hat{\bg} \times \dky \hat{\bg} \right)\,,
\end{align}
where $\hat{\bg}= {\bg}/|{\bg}|$, and the integration runs over the entire 2D Brillouin zone.
For the current model, we obtain
\begin{align}\label{eq:cnum2}
c(k_z) &= \frac{1}{2}[\mbox{sgn}(M_{\pi})-\mbox{sgn}(M_0)] \\
\mbox{with}\quad M_0 &= m_1(k_z) - m_2, \nonumber\\
M_{\pi} &= m_1(k_z) + m_2. \nonumber
\end{align}
$c(k_z)$ can only take the value $\pm1$ or $0$ \cite{Haldane:1988}. For a specific $k_z$, a non-zero $c$ characterizes a 2D QHI which accommodates one gapless chiral edge mode at its boundary (with the chirality given by the sign of $c$) ; while $c=0$ signals an NI for the effective 2D system with no surface states.
The value of $c(k_z)$ changes --accordingly
 a  topological phase transition happens for the effective 2D model-- only when the gap closes, that is, when $M_0(k_z) = 0$ at $(k_x,k_y) = (0,0)$,
 or/and when $M_{\pi}(k_z) = 0$ at $(k_x,k_y) = (\pi,\pi)$. Hence we recover exactly
 %\eqref{eq:zcond0} and \eqref{eq:zcondpi},
 the conditions \eqref{eq:zcond} for the Weyl points to occur in the 3D model.
 From this point of view,
 the Weyl points appear to be the critical points for the topological phase transitions in a 2D Haldane-like model,
 with $k_z$ playing the role of the controlling parameter.
It is also clear that within a certain range of $k_z$
where the reduced 2D model is topologically nontrivial ($\cos{k_z}<m_2-m_1$ for $m_1+m_2>1$), edge states persist
 and form gapless surface bands in 3D, while in the complementary range of $k_z$ (excluding the transition points),
 the reduced 2D model is topologically trivial and a (truly) gapped region is expected  --
the gapless surface bands cease to exist beyond the critical (Weyl) points therefore the Fermi arcs end at these spots. Furthermore, as the locations of the bulk Weyl points evolve,
the range of $k_z$ for which the (2D) topological order exists can either expand or shrink, resulting in eventually the QHI or the NI phase (see Fig.\ref{fig:surf}).

Next we treat both $k_x$ and $k_y$ as parameters and reduce the system effectively to a 1D chain along the $z$ axis.
We investigate the topological property associated with this 1D chain in order to account for the absence of nontrivial
surface bands for surfaces parallel to the $x$-$y$ plane. In this case, the topological invariant is the winding number
 of $\hat{\bg}$ --defined the same way as in Eq. \ref{eq:ham2D} for $(k_x,k_y) \ne (0,0)$ or $(\pi,\pi)$, but treated
 only as a function of $k_z$-- around a given axis when $k_z$ runs across the Brillouin zone (i.e. from $-\pi$ to $\pi$).
Without entering detailed calculations, one immediately finds this winding number to be identically zero regardless of
the values of $k_x$ and $k_y$, by simply noticing that $\hat{\bg}(k_z) = \hat{\bg}(-k_z)$, hence the trajectories of
$\hat{\bg}$ for $k_z \in [-\pi,0]$ and for $k_z \in [0,\pi]$ overlaps exactly apart from having opposite directions.
That is, the effective 1D chain considered in this case is always topologically trivial. Therefore no nontrivial boundary states, or nontrivial surface bands back to the 3D model, shall be expected for this specially chosen orientation.

\section{Global Phase Diagram}
\begin{figure}
    \centering
    \includegraphics[width=0.46\textwidth]{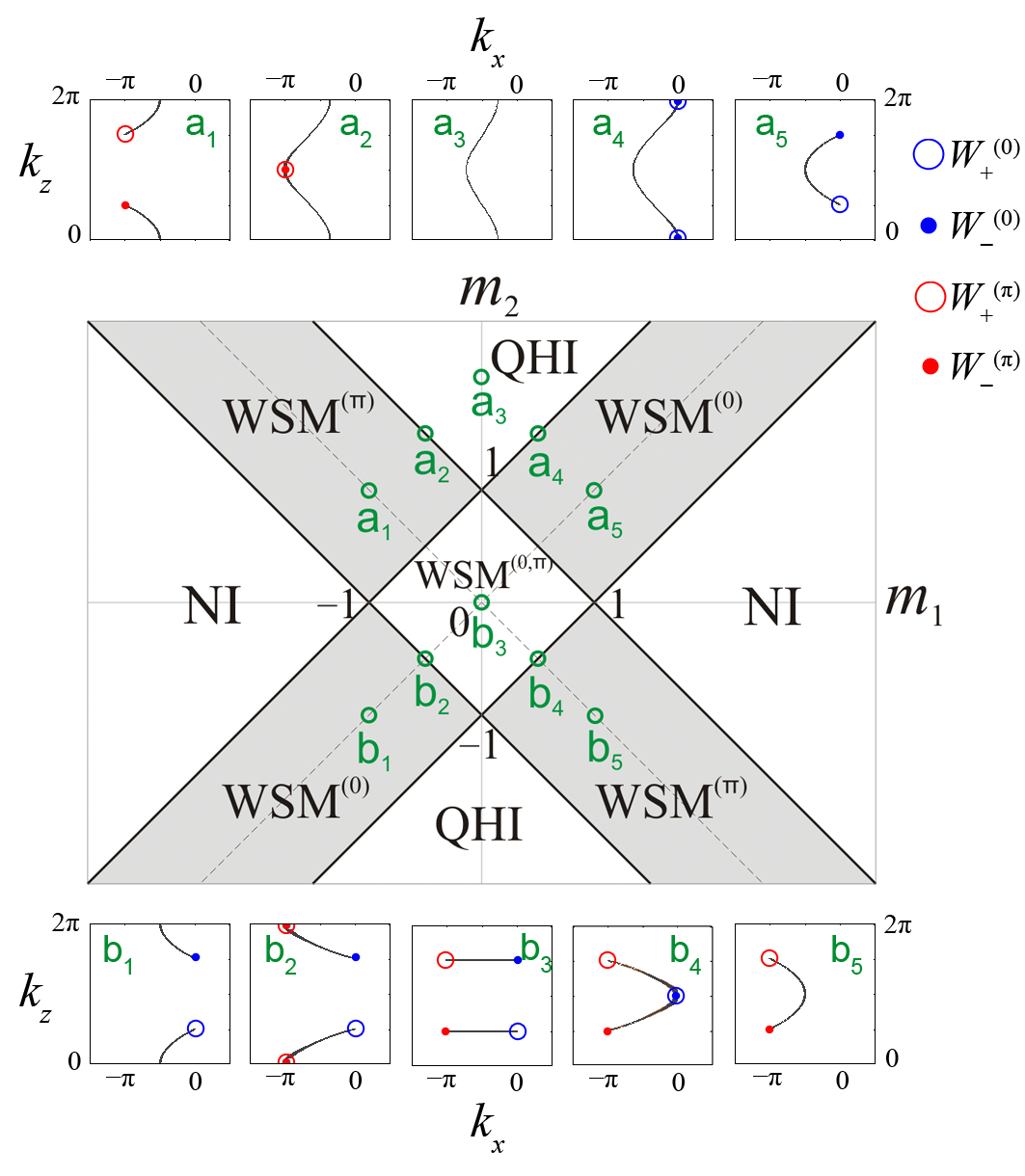}
    \caption{Full phase diagram of our model (the middle panel) and two selected trajectories along which the evolution of Fermi arcs is examined 
(the upper and the lower panels). In the phase diagram, the grey regions represent Weyl semi-metallic phases with only one pair of Weyl points either
 along $(k_x,k_y) = (0,0)$ (hence labeled WSM$^{(0)}$) or along $(k_x,k_y) = (\pi,\pi)$ (hence labeled WSM$^{(\pi)}$). Both pairs of Weyl points coexist
 in the region labeled WSM$^{(0,\pi)}$. The two types of bulk insulating phases, labeled QHI and NI, are topologically distinct in terms of the Chern 
numbers defined in the reduced dimensions, given by $c(k_z)=\pm1$ and $c(k_z)=0$ for every $k_z$, respectively. Along the trajectory (labeled \textbf{a}) 
passing through the QHI phase, Fermi arcs close before reopening, whereas along the trajectory (labeled \textbf{b}) passing through the WSM$^{(0,\pi)}$ phase, 
Fermi arcs divide before rejoining. The helicity of each Weyl point is represented by a circle ($+1$) or a dot ($-1$).}
    \label{fig:arcs}
\end{figure}
In the last part of this paper, we return to the full phase diagram with respect to $m_1=\Delta/2t_\perp$ and $m_2 = 2t'/t_\perp$,
which has been limited so far to the range $m_1,m_2>0$ and $m_1 + m_2>1$. We consider several symmetry operations
 to relate different quadrants of the phase diagram.
% In particular we constrain ourselves to symmetry operations on the bulk Hamiltonian \eqref{eq:ham3D}, which allows us to deduce, as we have shown previously, the main features of surface bands and Fermi arcs
% \footnote{On the other hand, the boundary conditions, which determine the specific details of surface bands and Fermi arcs,
%are not included in this symmetry consideration.}.
Explicitly, in order to
%map two mirror image points with respect to the $m_2$-axis (i.e.
relate two points $(m_1,m_2)$ and $(-m_1,m_2)$ in the phase diagram,
it is sufficient to transform $(k_x,k_y,k_z) \rightarrow (k_x,k_y,k_z)+(\pi,\pi,\pi)$ and
$\ham \rightarrow \sigma_y \ham \sigma_y$; in order to map
%two mirror image points with respect to the $m_1$-axis (i.e.
 $(m_1,m_2)$ to $(m_1,-m_2)$, we use the transformation $(k_x,k_y,k_z) \rightarrow (k_x,k_y,k_z)+(\pi,\pi,0)$
and $\sigma_x \rightarrow -\sigma_x$; combined together, the transformations $(k_x,k_y,k_z) \rightarrow (k_x,k_y,k_z)+(0,0,\pi)$
and $\sigma_z \rightarrow -\sigma_z$ map $(m_1,m_2)$ to $(-m_1,-m_2)$.

These symmetry operations immediately extend our knowledge of the bulk bands from the first quadrant of the phase
diagram to the rest (see the middle panel of Fig. \ref{fig:arcs}) \footnote{On the other hand, the boundary conditions, which determine the specific details of surface bands and Fermi arcs, are not included in this symmetry consideration.}. Namely, the two individual regions given by $|m_1-m_2|<1$ and
$|m_1+m_2|>1$ are WSM regimes with the two Weyl points occurring along $(k_x,k_y)=(0,0)$, hence are labeled WSM$^{(0)}$; the two individual regions given by $|m_1+m_2|<1$ and $|m_2-m_1|>1$ are also WSM regimes but with the two Weyl points occurring along $(k_x,k_y)=(\pi,\pi)$, hence are labeled WSM$^{(\pi)}$; the semi-metal phase labeled WSM$^{(0,\pi)}$ possesses two pairs of Weyl points, occurring along $(k_x,k_y)=(0,0)$ and $(k_x,k_y)=(\pi,\pi)$; the previously identified QHI and NI regimes are now extended by symmetry.

In this phase diagram, it is instructive to examine the evolution of Fermi arcs and their underlying surface bands in a coherent manner. This evolution is presented in Fig.~\ref{fig:arcs} for a surface parallel to the $x$-$z$ plane and for two different trajectories labeled \textbf{a} and \textbf{b} (see Fig. \ref{fig:arcs}). Along trajectory \textbf{a}, the open Fermi arc in the WSM$^{(0)}$ phase closes before opening differently in the WSM$^{(\pi)}$ phase when passing through the QHI phase (see the upper panel of Fig. \ref{fig:arcs}); along trajectory \textbf{b}, the open Fermi arc in the WSM$^{(0)}$ phase divides into two pieces before joining differently in the WSM$^{(\pi)}$ phase when passing through the WSM$^{(0,\pi)}$ phase \footnote{The Fermi arc and surface band features of the WSM$^{(0,\pi)}$ phase can also be analyzed by using the dimensional reduction method, of which we will not enter details in this paper.} (see the lower panel of Fig. \ref{fig:arcs}).
%
%It is instructive to compare the two trajectories:
%both trajectories run from one of the WSM$^{(0)}$ and WSM$^{(\pi)}$ phases to the other, but trajectory \textbf{a} passes the I1 phase
% whereas trajectory \textbf{b} passes the WSM$\times$2 phase (see the middle panel of Fig. \ref{fig:arcs}). Phenomenologically, the I1 phase,
%containing no Weyl point but a closed Fermi arc, bridges two WSM phases by ``cutting" its Fermi arc differently (see the upper panels of
%Fig. \ref{fig:arcs}),
%while the WSM$\times$2 phase, containing four Weyl points and two disconnected Fermi arcs \footnote{The Fermi arc and
% surface band features of the WSM$\times$2 phase can also be analyzed by using the dimensional reduction method, of which we will not enter
%details in this paper.}, bridges two WSM phases by ``joining" its Fermi arcs differently (see the lower panels of Fig. \ref{fig:arcs}).
%
Another notable feature in the full phase diagram is the reversing of the surface band chirality (with respect to the $z$ axis) and the associated sign change of $c(k_{z})$ across the $m_2=0$ line. Along the $m_2=0$ line, the allowed Fermi arcs appear parallel to the $k_x$ axis and the underlying surface bands are {\it dispersionless} in the $x$ direction. Indeed, the sign of $m_2$ encodes the direction of the magnetic flux penetrating through the $x$-$y$ plane in our original lattice model.
%As a direct consequence, the topological number $c$ has to be
%$0$ in the insulating phases I2 that are delimited by semi-metal phases with opposite chirality.

In this paper, we have proposed a simple tight-binding model of spinless fermions which gives rise to a 3D topological Weyl semi-metal phase in its phase diagram. We have investigated in detail this Weyl semi-metal phase and its related insulating phases, especially in terms of their topological properties and the resulting surface bands. In particular, we have shown that the Weyl semi-metal occurs between a $d=2+1$ quantum Hall and a normal insulator, in a manner similar to the semi-metal occurring between a $\mathbb{Z}_{2}$ topological and a normal insulator \cite{Murakami:2007}. We believe that beyond its intrinsic interests, our lattice model is a perfect test bench for various properties of the Weyl semi-metal phase proposed in various materials, allowing in particular for numerical studies of its transport properties.

Upon finishing this work we became aware of a preprint which focuses on a different lattice model for Weyl fermions \cite{Jiang:2011}.

\acknowledgments
 D.C. acknowledges ANR support through grant BLANC-10 IsoTop. In Geneva P. D. was supported by the European Marie Curie ITN  NanoCTM and J. L. was supported by the Swiss National Science Foundation and the Swiss National Center of Competence in Research on Quantum Science and Technology.

\bibliographystyle{eplbib}

\bibliography{weyl3d_refs}

\end{document}